\documentclass{elsart}
\journal{Physica A}
\newcommand \be{ \begin{equation}}
\newcommand \ee{ \end{equation} }
\newcommand \bea{ \begin{eqnarray} }
\newcommand \eea{ \end{eqnarray} }
\newcommand \nn{ \nonumber }
\newcommand \ph{ \phi }
\newcommand \pfa{ {\rm Pf} } 
\newcommand \bra{ \langle }
\newcommand \ket{ \rangle }
\newcommand \nd{{\vphantom{\dagger}}} 
\newcommand \idmat{ I } 
\newcommand \im{ {i} } 
\newcommand \dif{ d } 
\newcommand \detB {\mathcal{B}} 
\newcommand \detC {\mathcal{C}} 
\newcommand \detM {\mathcal{M}} 
\newcommand \vecr {{\bf r}} 
\newcommand \m { $b$ }
\newcommand \p { $a$ }
\newcommand \id {s}
\newcommand \s {\sigma}
\newcommand \fig[1]{Fig.~\ref{#1}}

\newcommand \sect[1]{Sec.~\ref{#1}}
\newcommand \citejournal[4]{{\it #1}, {\bf #2} (#4), p.#3 } 

\newcommand \paperI{I}

\usepackage{graphicx}
\usepackage{amsmath}

\begin{document}
\begin{frontmatter}
\title{Dimers on the kagome lattice I: 
Finite lattices}

\author[Boston]{F. Y. Wu},
\author[Berkeley,LBL]{Fa Wang}
\address[Boston]{ Department of Physics, Northeastern University, Boston, Massachusetts 02115 }
\address[Berkeley]{ Department of Physics, University of California, Berkeley, California 94720 }
\address[LBL]{ Material Sciences Division, Lawrence Berkeley National Laboratory, Berkeley, California 94720 }

\date{Printed \today}

\begin{abstract}
We report exact results on the
enumeration of close-packed dimers on a {\it finite} kagome lattice
with general  {\it asymmetric} dimer weights under periodic and cylindrical boundary conditions.
For symmetric dimer weights, the resulting dimer generating functions 
reduce to very simple expressions, and we show how the simple expressions can
be obtained from the consideration of a spin-variable mapping.
  
\end{abstract}


\end{frontmatter}

\section{Introduction}\label{sec:introduction}
A central problem in lattice statistics is the enumeration of
close-packed dimers on lattices and graphs. The origin of the
problem has a long history dating back to a 1937 paper by Fowler
and Rushbrooke  \cite{Fowler} in an attempt of enumerating the absorption of
diatomic molecules on a surface. A breakthrough in
dimer statistics has been the exact solution of the generating function
for a finite square lattice of size $M\times N$, where $M$ and $N$ are arbitrary,
 obtained by Kasteleyn
\cite{Kasteleyn:physica} and by Temperley and Fisher
\cite{Temperley} in 1961.

 In view of the role of  finite-size solutions
 in the conformal field theory discovered by Bl\"ote {\it et al.} 
\cite{cardy} in 1986, it has been of increasing importance to consider solutions of lattice models
 for various finite
two-dimensional  lattices. Thus, the dimer solution has been extended
to cylindrical \cite{McCoyWu} and nonorientable \cite{LuWu} lattices.
However, these lattices are variants of the square lattice
which may not necessarily exhibit special lattice-dependence features.

In a recent paper \cite{WangWu} we have reported  enumeration results
of close-packed dimers on an infinite kagome lattice with symmetric dimer
weights (activities). The solution turned out to assume a very simple
expression. 
In this paper we extend the solution to {\it finite} lattices with  general
{\it asymmetric} weights and
 under two different boundary conditions. We find the solutions  given by entirely
different expressions. For symmetric weights, however, the solutions
  again reduce to simple expressions. We 
  show how the simple expressions can be deduced
 quite directly from the consideration of a spin-variable mapping.
 
\section{Finite lattices}\label{sec:finite}
\begin{figure}
\includegraphics{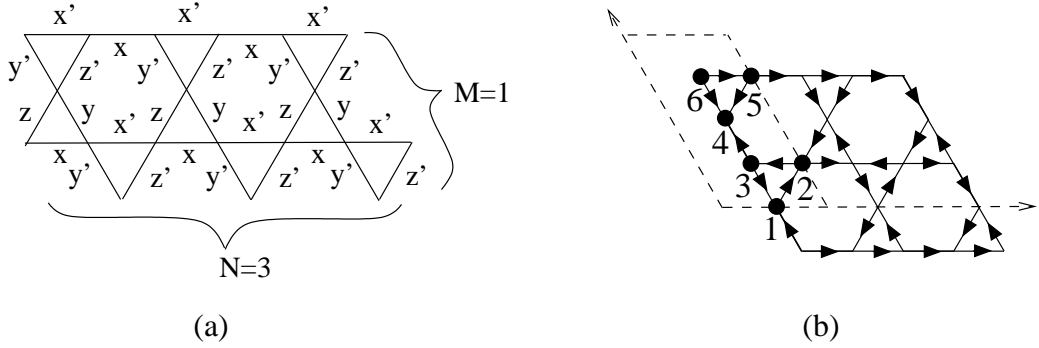}
\caption{(a) An $M\times N$ lattice of $6MN$ sites with asymmetric
dimer weights. 
(b) A Kasteleyn edge orientation of the kagome lattice. A
unit cell is the region bounded by broken lines.}
\label{fig:kagome}
\end{figure}

Consider a kagome lattice with {asymmetric} dimer weights
$x,y,z$ around up-pointing triangles and $x',y',z'$ around
down-pointing triangles as shown in \fig{fig:kagome}(a). 
We
consider a lattice of $M\times N$ unit cells having a total of
$6MN$ lattice sites; the case of $M=1,N=3$ is shown
in \fig{fig:kagome}(a). A unit cell
of the lattice
contains $6$ sites numbered $1, 2, ..., 6$ as indicated in
\fig{fig:kagome}(b). The  Kasteleyn edge orientations
adopted in
\cite{WangWu} is also shown in \fig{fig:kagome}(b).
 
The method of Pfaffians \cite{Kasteleyn:physica} concerns with the evaluation of 
a $6MN\times 6MN$ antisymmetric {\it Kasteleyn matrix} $A$ written down according to 
edge weights and orientations (under specific boundary conditions) which can be
read off from 
Figs~\ref{fig:kagome}(a) and \ref{fig:kagome}(b),
and by adopting the prescription
 \be
 A_{ij}=\left\{\begin{array}{lr}
 +w_{ij} & {\rm orientation\  from\ }i{\rm \ to\ }j\\
 -w_{ij} & {\rm orientation\  from\ }j{\rm \ to\ }i
 \end{array}\right.
 \label{equ:Aij} 
 \ee 
where $w_{ij}$ is the weight of edge $ij$.
The dimer generating function is then given by the square root of the
determinant of the matrix $A$.
We consider the kagome lattice under two different boundary conditions.

 \subsection{The periodic boundary condition}\label{sec:periodic}
First we consider the periodic (toroidal) boundary condition (PBC) for which the
lattice is periodic in both the horizontal and vertical
directions.
Kasteleyn
\cite{Kasteleyn:physica} has shown that under the PBC the dimer
generating function $Z_{\rm PBC}$ is a linear combination of four
Pfaffians Pf$|A_i|, i=1,2,3,4$,
 \be
 Z_{\rm PBC}=\frac{1}{2}\Big[-{\rm Pf}|A_1|+{\rm Pf}|A_2|
  +{\rm Pf}|A_3|+{\rm Pf}|A_4|\Big]. \label{PBCGen1}
 \ee
Up to signs yet to be determined, the Pfaffians are the square
root of the determinants specified by the Kasteleyn orientation
of lattice edges with, or without, the reversal of  arrows on  edges
connecting two opposite boundaries. A perusal of \fig{fig:kagome}(b)
and the use of the prescription (\ref{equ:Aij}) 
lead to the four $6MN\times 6MN$ Kasteleyn matrices,
  \be
\begin{split}
  A_1  = & a_{0,0}\otimes \idmat_M\otimes \idmat_N
    + a_{1,0}    \otimes \idmat_M\otimes T_N
    - a_{1,0}^{T}\otimes \idmat_M\otimes T_N^{T} \\
 & + a_{0,1}    \otimes T_M     \otimes \idmat_N
    - a_{0,1}^{T}\otimes T_M     \otimes \idmat_N^{T} \\
 & + a_{1,1}    \otimes T_M     \otimes T_N
    - a_{1,1}^{T}\otimes T_M^{T} \otimes T_N^{T} \\
A_2  = & a_{0,0}\otimes \idmat_M\otimes \idmat_N
    + a_{1,0}    \otimes \idmat_M\otimes H_N
    - a_{1,0}^{T}\otimes \idmat_M\otimes H_N^{T} \\
 & + a_{0,1}    \otimes T_M     \otimes \idmat_N
    - a_{0,1}^{T}\otimes T_M     \otimes \idmat_N^{T} \\
 & + a_{1,1}    \otimes T_M     \otimes H_N
    - a_{1,1}^{T}\otimes T_M^{T} \otimes H_N^{T} \\
A_3  = & a_{0,0}\otimes \idmat_M\otimes \idmat_N
    + a_{1,0}    \otimes \idmat_M\otimes T_N
    - a_{1,0}^{T}\otimes \idmat_M\otimes T_N^{T} \\
 & + a_{0,1}    \otimes H_M     \otimes \idmat_N
    - a_{0,1}^{T}\otimes H_M     \otimes \idmat_N^{T} \\
 & + a_{1,1}    \otimes H_M     \otimes T_N
    - a_{1,1}^{T}\otimes H_M^{T} \otimes T_N^{T} \\
A_4  = & a_{0,0}\otimes \idmat_M\otimes \idmat_N
    + a_{1,0}    \otimes \idmat_M\otimes H_N
    - a_{1,0}^{T}\otimes \idmat_M\otimes H_N^{T} \\
 & + a_{0,1}    \otimes H_M     \otimes \idmat_N
    - a_{0,1}^{T}\otimes H_M     \otimes \idmat_N^{T} \\
 & + a_{1,1}    \otimes H_M     \otimes H_N
    - a_{1,1}^{T}\otimes H_M^{T} \otimes H_N^{T}. 
\end{split}
\label{equ:Athph}
 \ee

Here, 
the superscripts $T$ denote  transpose,  $\otimes$ is direct product,
 $\idmat_M$ is the $M\times M$ identity matrix, and $H_N$,
$T_N$ are  the $N\times N$ matrices
\be
\begin{split}
H_N=\begin{pmatrix}
 0     & 1     & 0     &\cdots & 0 \\
 0     & 0     & 1     &\cdots & 0 \\
\vdots &\vdots &\vdots &\ddots &\vdots \\
 0     & 0     & 0     &\cdots & 1 \\
-1     & 0     & 0     &\cdots & 0
\end{pmatrix} , \quad
T_N=\begin{pmatrix}
 0     & 1     & 0     &\cdots & 0 \\
 0     & 0     & 1     &\cdots & 0 \\
\vdots &\vdots &\vdots &\ddots &\vdots \\
 0     & 0     & 0     &\cdots & 1 \\
 1     & 0     & 0     &\cdots & 0
\end{pmatrix}, \quad 
F_N=\begin{pmatrix}
 0     & 1     & 0     &\cdots & 0 \\
 0     & 0     & 1     &\cdots & 0 \\
\vdots &\vdots &\vdots &\ddots &\vdots \\
 0     & 0     & 0     &\cdots & 1 \\
 0     & 0     & 0     &\cdots & 0
\end{pmatrix}, \label{HTF}
 \end{split}
 \ee
where $F_N$ is to be used later in (\ref{ACBC}), and
 \bea
 a_{0,0} & = & \begin{pmatrix}
 0 &  z'& -y'&  0 &  0 &  0 \cr
-z'&  0 &  x'&  0 &  0 &  0 \cr
 y & -x'&  0 &  y &  0 &  0 \cr
 0 &  0 & -y &  0 & -z'& -y'\cr
 0 &  0 &  0 &  z'&  0 & -x'\cr
 0 &  0 &  0 &  y'&  x'&  0 \cr
\end{pmatrix},\quad
a_{1,0} = \begin{pmatrix}
 0 &  0 &  0 &  0 &  0 &  0 \cr
 0 &  0 &  x & -z &  0 &  0 \cr
 0 &  0 &  0 &  0 &  0 &  0 \cr
 0 &  0 &  0 &  0 &  0 &  0 \cr
 0 &  0 &  0 &  0 &  0 &  x \cr
 0 &  0 &  0 &  0 &  0 &  0 \cr
\end{pmatrix}, \nn\\
a_{0,1} & = & \begin{pmatrix}
 0 &  0 &  0 &  0 &  0 &  0 \cr
 0 &  0 &  0 &  0 &  0 &  0 \cr
 0 &  0 &  0 &  0 &  0 &  0 \cr
 0 &  0 &  0 &  0 &  0 &  0 \cr
 0 &  0 &  0 &  0 &  0 &  0 \cr
 y &  0 &  0 &  0 &  0 &  0 \cr
\end{pmatrix} ,\quad 
a_{1,1}   =  \begin{pmatrix}
 0 &  0 &  0 &  0 &  0 &  0 \cr
 0 &  0 &  0 &  0 &  0 &  0 \cr
 0 &  0 &  0 &  0 &  0 &  0 \cr
 0 &  0 &  0 &  0 &  0 &  0 \cr
-z &  0 &  0 &  0 &  0 &  0 \cr
 0 &  0 &  0 &  0 &  0 &  0 \cr
\end{pmatrix}, \label{equ:amatrices}\\  \nn  \\
a_{-1,0} &=& - a_{1,0}^{T} ,\quad a_{0,-1}   =   -
a_{0,1}^{T},\quad a_{-1,-1}   =   - a_{1,1}^{T}. \nn
 \eea

The determinant of a matrix is equal to the product
of its eigenvalues. To determine eigenvalues of
$A_{1,2,3,4}$, we first  block-diagonalize the 4 matrices
by appropriate Fourier transforms.
Since $T_N$ and $T_N^{T}$ commute, they can be simultaneously
diagonalized and replaced by respective eigenvalues $e^{\im\theta_n}$
and $e^{-\im\theta_n}$, where $\theta_n=2\pi n/N,\ n=0,\dots, N-1$.

Similarly, $T_M$ and $T_M^{T}$ can be simultaneously diagonalized
and replaced by eigenvalues $e^{\im\ph_m}$ and $e^{-\im\ph_m}$,
where $\ph_m=2\pi m/M,\ m=0,\dots, M-1$.

Likewise, $H_N$ and $H_N^{T}$ commute and they can be
simultaneously diagonalized and replaced by eigenvalues
$e^{\im\alpha_n}$ and $e^{-\im\alpha_n}$, where
$\alpha_n=(2n+1)\pi /N,\ n=0,\dots, N-1$; $H_M$ and $H_M^{T}$
can be simultaneously diagonalized and replaced by eigenvalues
$e^{\im\beta_m}$ and $e^{-\im\beta_m}$, where $\beta_m=(2m+1)\pi
/M,\ m=0,\dots, M-1$. Then we find
 \be
\begin{split}
&\det|A_1|  = \prod_{n=0}^{N-1}\prod_{m=0}^{M-1}
\det|A(\theta_n,\ph_m)|, \qquad
\det|A_2|  = \prod_{n=0}^{N-1}\prod_{m=0}^{M-1}
\det|A(\alpha_n,\ph_m)|, \\
&\det|A_3|  = \prod_{n=0}^{N-1}\prod_{m=0}^{M-1}
\det|A(\theta_n,\beta_m)|, \qquad
\det|A_4|  = \prod_{n=0}^{N-1}\prod_{m=0}^{M-1}
\det|A(\alpha_n,\beta_m)|,
\end{split}
\label{Amatrix}
 \ee
where the $6\times 6$ matrix $A(\theta,\ph)$ is anti-hermitian and is
given by
\bea
A(\theta,\ph) & = & a_{0,0} + a_{1,0} e^{\im\theta} + a_{-1,0} e^{-\im\theta}
 + a_{0,1} e^{\im\ph} + a_{0,-1} e^{-\im\ph} + a_{1,1} e^{\im(\theta+\ph)} \nn\\
&& + a_{-1,-1} e^{-\im(\theta+\ph)}\nn \\
& = & \begin{pmatrix}
 0                  &  z'                & -y'              &  0            &  z e^{-\im(\theta+\ph)} & -y e^{-\im\ph}\cr
-z'                 &  0                 &  x' +xe^{\im\theta} & -z e^{\im\theta} &  0                   &  0\cr
 y                  & -x' -xe^{-\im\theta}  &  0               &  y            &  0                   &  0\cr
 0                  &  z e^{-\im\theta}     & -y               &  0            & -z'                  & -y'\cr
-z e^{\im(\theta+\ph)} &  0                 &  0               &  z'           &  0                   &  -x' + xe^{\im\theta}\cr
 y e^{\im\ph}       &  0                 &  0               &  y'           &  x'-xe^{-\im\theta}     &  0\cr
\end{pmatrix} \nn\\
\label{equ:Aphitheta}
\eea
This yields
 \be
   \det A(\theta, \ph) = 2A +2D\cos(\ph) + 2E\cos(2\theta+\ph)
               +4\Delta_2 \sin^2\theta  \nn
               \ee
with
 \be
\begin{array}{rcl}
&&A \ =  (x y' z+x' y z')^2+(x' y' z+x y z')^2\\
&&D \ =  -(x y' z-x' y z')^2\\
&&E \ \,=  (x' y' z-x y z')^2\\
&&\Delta_2  =  (x' y z-x y' z')^2.
\end{array}
\label{Delta}
 \ee
 The desired generating function is now obtained
 by substituting either Pf$|A_i| = + \sqrt{ \det|A_i|}$
 or $- \sqrt{ \det|A_i|}$ into
 (\ref{PBCGen1}).  In the present case
 the signs in front  of the square roots can be determined by considering
  the  case of $M=N=1$.  By explicit enumeration we have
 \be
 Z_{\rm PBC} = 2 (x+x')(yz'+y'z), \qquad M=N=1. \label{11}
 \ee
It is readily verified that the expression (\ref{11}) is
reproduced by (\ref{PBCGen1}) if all 4 terms in (\ref{PBCGen1})
are positive. Thus, we are led to the final expression
 \be
 Z_{\rm PBC}=\frac{1}{2}\Big[\sqrt{\det|A_1|}+\sqrt{\det|A_2|}
  +\sqrt{\det|A_3|}+\sqrt{\det|A_4|}\Big], \label{PBCGen}
 \ee
where $\det |A_i|$, $i=1, ...,4$, are given by (\ref{Amatrix}).

For symmetric dimer weights $x'=x,y'=y,z'=z$, we have
$D=E=\Delta_2=0$, $\det|A_{1,2,3,4}|=(4 x y z)^{MN}$, and the
simple result
 \be
 Z_{\rm PBC}=2\cdot(4 x y z)^{MN}, \quad x'=x,\ y'=y,\ z'=z.
 \label{Asym}
 \ee
We shall see in \sect{sec:spin}  that this simple result can be
understood and deduced  directly  using a spin-variable mapping.

In the case of an infinite lattice, (\ref{PBCGen1}) leads to the
per-dimer free energy
 \bea
 f &=& \lim_{M,N \to\infty} \frac 1 {3MN} \ln Z_{\rm PBC} \nn \\
 &=& \frac{1}{24\pi^2}\int_{0}^{2\pi}  \dif\theta
\int_{0}^{2\pi}  \dif\ph \ln
\Big[2A+2D\cos(\theta-\ph) \nn \\
&& \hskip2cm +2E\cos(\theta+\ph)+4\Delta_2 \sin^2\theta \Big].
 \label{freePBC}
 \eea
This free energy is independent of the boundary condition.
 The free energy
(\ref{freePBC})
 can also be deduced using the vertex-model approach introduced
 in \cite{WangWu}, details of which are straightforward and will
 not be given.

For symmetric dimer weights $x'=x, y'=y, z'=z$, (\ref{freePBC}) reduces 
further to
\be
f = \frac 1 3 \ln (4xyz). \label{freePBC1}
\ee
This result for an infinite lattice was first reported in \cite{wureview}
with the full derivation given  in \cite{WangWu}. The exact per-dimer entropy  
$s = \frac 2 3 \ln 2 $\, obtained
from (\ref{freePBC1})
at $x=y=z=1$ has been cited earlier by Phares and Wunderlich 
from  \cite{phares} and by Elser \cite{elser} from 
different considerations. 
 
\subsection{The cylindrical boundary condition}\label{sec:cylindrical}
Consider next the  cylindrical boundary condition (CBC) for which
the lattice of $M\times N$ unit cells  is periodic in the
horizontal  direction. The Kasteleyn orientation is achieved by
reversing the orientations of the $4M-1$ edges connecting unit
cells in the $ N $th column to those in the first column.
This gives
the dimer generating function as a single Pfaffian
 \be
Z_{\rm CBC}=\sqrt{\det|A_{\rm CBC}|},\label{cbcz}
 \ee
where $ A_{\rm CBC} $ is the $ 6MN\times 6MN $
matrix
\be
\begin{split}
A_{\rm CBC}  = & a_{0,0}\otimes \idmat_M\otimes \idmat_N + a_{1,0}    \otimes \idmat_M\otimes H_N
 - a_{1,0}^{T}\otimes \idmat_M\otimes H_N^{T} + a_{0,1}    \otimes F_M \otimes \idmat_N \\
& - a_{0,1}^{T}\otimes F_M    \otimes \idmat_N^{T} + a_{1,1}    \otimes F_M \otimes H_N - a_{1,1}^{T}\otimes F_M^{T}\otimes H_N^{T}. \label{ACBC}
\end{split}
 \ee
Here, matrices  $ a $ are  those   in (\ref{equ:amatrices}) and
$F_N $ has been given in (\ref{HTF}).
Again, the matrix $A_{\rm CBC}$ is block-diagonalized by replacing $H_N$ and
$H_N^T$ by their respective eigenvalues. This leads to
 \be
 \det|A_{\rm CBC}|=\prod_{n=0}^{N-1}{\det|B_M(\theta_n)|}, \quad \quad \theta_n =
 (2n+1)\pi/N
\label{prodB}
 \ee
 where $B_M(\theta)$ is the
$6M\times 6M$ Kasteleyn matrix
  \be
\begin{split}
B_M(\theta) & = B\otimes \idmat_M + B_{+}\otimes F_M + B_{-}\otimes F_M^{T} \\
& = \begin{pmatrix}
 B     & B_{+} & 0     &\cdots & 0     & 0     & 0 \\
 B_{-} & B     & B_{+} &\cdots & 0     & 0     & 0 \\
 0     & B_{-} & B     &\cdots & 0     & 0     & 0 \\
\vdots &\vdots &\vdots &\ddots &\vdots &\vdots &\vdots \\
 0     & 0     & 0     &\cdots & B     & B_{+} & 0 \\
 0     & 0     & 0     &\cdots & B_{-} & B     & B_{+} \\
 0     & 0     & 0     &\cdots & 0     & B_{-} & B
\end{pmatrix},
\end{split}
 \ee
 $B = a_{0,0}+e^{\im\theta}a_{1,0}-e^{-\im\theta}a_{1,0}^{T}$, $B_{+} =a_{0,1}+ e^{ \im\theta}a_{1,1} $, and
 $B_{-} =-a_{0,1}^T- e^{-\im\theta}a_{1,1}^T $ are the $6\times 6$ matrices
 \bea
  B& = & 
\begin{pmatrix}
 0 &  z'               & -y'               &  0             &  0                &  0  \\
-z'&  0                &  x'+xe^{ \im\theta}  & -z e^{ \im\theta} &  0                &  0  \\
 y'& -x'-xe^{-\im\theta}  &  0                &  y             &  0                &  0  \\
 0 &  z e^{-\im\theta}    & -y                &  0             & -z'               & -y' \\
 0 &  0                &  0                &  z'            &  0                & -x'+xe^{ \im\theta} \\
 0 &  0                &  0                &  y'            &  x'-xe^{-\im\theta} &  0
\end{pmatrix},
\label{b} \\
 B_{+} &=&
\begin{pmatrix}
  0             & 0 & 0 & 0 & 0 & 0 \\
  0             & 0 & 0 & 0 & 0 & 0 \\
  0             & 0 & 0 & 0 & 0 & 0 \\
  0             & 0 & 0 & 0 & 0 & 0 \\
 -z e^{ \im\theta} & 0 & 0 & 0 & 0 & 0 \\
  y             & 0 & 0 & 0 & 0 & 0
\end{pmatrix},    \qquad \qquad
B_{-} =
\begin{pmatrix}
 0 & 0 & 0 & 0 & z e^{-\im\theta} & -y \\
 0 & 0 & 0 & 0 & 0             &  0 \\
 0 & 0 & 0 & 0 & 0             &  0 \\
 0 & 0 & 0 & 0 & 0             &  0 \\
 0 & 0 & 0 & 0 & 0             &  0 \\
 0 & 0 & 0 & 0 & 0             &  0
\end{pmatrix}.\nn
 \eea
 The matrix $B_M(\theta)$ is of  a form of that occurring in
 the evaluation  of an Ising partition function under the cylindrical boundary
 condition \cite{McCoyWu,LuWu}, and the determinant $\det|B_M(\theta)|$ can be  evaluated
 as follows:

   Let
$B^{i,j}$ denote the $6\times 6$ matrix $B$ with row $i$ and
column $j$  removed. By Laplacian expansion  and the use of a
lemma established in \cite{LuWu}, the determinant of $B_M$ can be
expanded as
  \bea
\det|B_M| &=&  \det|B| \cdot \det|B_{M-1}|
  + z^2 \det|B^{5,5}| \cdot \det|B_{M-1}^{1,1}| \nn \\
&& + y^2 \det|B^{6,6}| \cdot \det|B_{M-1}^{1,1}| + y z e^{ \im\theta} \det|B^{5,6}| \cdot \det|B_{M-1}^{1,1}| \nn \\
&&  + y z e^{-\im\theta} \det|B^{6,5}| \cdot \det|B_{M-1}^{1,1}|.
   \label{recu1}
 \eea
Similarly, the determinant of the matrix $B_M^{1,1}$ can be expanded as
  \bea
\det|B_M^{1,1}|& = &  \det|B^{1,1}| \cdot \det|B_{M-1}|
  + z^2 \det|B^{1,1;5,5}| \cdot \det|B_{M-1}^{1,1}| \nn \\
&&  + y^2 \det|B^{1,1;6,6}| \cdot \det|B_{M-1}^{1,1}| + y z e^{ \im\theta} \det|B^{1,1;5,6}| \cdot \det|B_{M-1}^{1,1}| \nn \\
&&  + y z e^{-\im\theta} \det|B^{1,1;6,5}| \cdot  \det|B_{M-1}^{1,1}|,
  \label{recu2}
 \eea
 where $B^{i,j;k,\ell}$ is the $6\times 6$ matrix $B$ with
 rows $i$ and $k$, and columns $j$ and $\ell$ deleted.

Write $\detB_M\equiv \det|B_M|$ and $\detC_M \equiv
\det|B_M^{1,1}|$. Expansions (\ref{recu1}) and (\ref{recu2}) are
recursion relations of $\detB$ and $\detC$,
 \bea
\detB_M & = & a\,\detB_{M-1} + b\,\detC_{M-1} \nn \\
\detC_M & = & c\,\detB_{M-1} + d\,\detC_{M-1}  \label{recuBC}
 \eea
 where
  \be
\begin{split}
a  = & \det |B| \\
b  = & z^2\det|B^{5,5}| + y^2 \det|B^{6,6}| + y z e^{ \im\theta} \det|B^{5,6}| + y z e^{-\im\theta} \det|B^{6,5}|  \\
c  = & \det|B^{1,1}| \\
d  = & z^2\det|B^{1,1;5,5}| + y^2 \det|B^{1,1;6,6}| + y z e^{ \im\theta} \det|B^{1,1;5,6}| + y z e^{-\im\theta} \det|B^{1,1;6,5}|,
\end{split}
 \label{abcd}
 \ee
subject to the  initial condition  $\detB_0=1$, $\detC_0=0$.
Explicitly using (\ref{b}), (\ref{abcd}) reads
 \be
\begin{split}
a  = & (x^2+x'^2)(y'^2 z^2 + y^2 z'^2) + 4 x^2 y'^2 z'^2 - 2(x' y' z + x y z') (x y' z + x' y z')\cos\theta \\
& + 4 x y' z'(x' y z - x y' z')\cos^2\theta \\
 b  = & 2\im (x' y z + x y' z')(y'^2 z^2 + y^2 z'^2)\sin\theta - 4\im y y' z z'(x' y z - x y' z')\cos\theta \sin\theta  \\
 c  = & -2\im \Big[(x^2 + x'^2)(x' y z + x y' z') - 2 x x'(x' y z - x y' z')\cos\theta\Big ]\sin\theta \\
d  = & (x^2+x'^2)(y'^2 z^2 + y^2 z'^2)  + 4 x'^2 y^2 z^2  + 2(x' y' z + x y z') (x y' z + x' y z')\cos\theta \\
& - 4 x' y z(x' y z - x y' z')\cos^2\theta .
\end{split}
 \ee

The recursion
 relation (\ref{recuBC}) can be solved by introducing
 generating functions
 \be
 {\bf B}( t )=\sum_{M=0}^{\infty}{\detB_M  t  ^M},\quad {\bf C}( t )=\sum_{M=0}^{\infty}{\detC_M  t  ^M}. \label{BCgen}
 \ee
 The recursion
relation (\ref{recuBC}) gives
 \bea
{\bf B}( t  ) & = &1+ t  \,[a\,{\bf B}( t  ) + b\,{\bf C}( t  )] \nn \\
{\bf C}( t  ) & = &  t  \,[c\,{\bf B}( t  ) + d\,{\bf C}( t  )],
\label{BC}
 \eea
where we have made  use of the initial condition $\detB_0=1$, $\detC_0=0$.

Solving (\ref{BC}) for ${\bf B}(t)$ and ${\bf C}(t)$, we obtain
  \be {\bf B}( t  )=
   \frac{1-d\, t  }{1-(a+d) t  +(a\,d - b\,c) t  ^2} = \frac
{1-d\,t} {(1-\lambda_+ t)(1-\lambda_- t)}, \label{Bt}
 \ee
where
\be
 \begin{split}
 \lambda_{\pm}= &(a+d)/2\pm\sqrt{(a-d)^2 / 4 + b c}  \\
 = & A+2\Delta_2 \sin^2\theta \pm \sqrt{ (A+2\Delta_2 \sin^2\theta)^2 -D^2 -E^2 -2DE \cos^2 \theta}.
 \label{lambda}
 \end{split}
\ee

 Here,  $A,D,E,\Delta_2$ have been given in (\ref{Delta}).
 Partial fraction and expand the right-hand side of (\ref{Bt}), and
 compare the resulting expansion with (\ref{BCgen}), one obtains
\be
 \detB_M(\theta)=\frac{\lambda_+^{M+1}-\lambda_-^{M+1}}{\lambda_+
 -\lambda_-}-\frac{\lambda_+^{M}-\lambda_-^{M}}{\lambda_+-\lambda_-}\cdot d.  
 \label{detNM}
\ee
 Finally, by combining (\ref{cbcz}) and (\ref{prodB}), we obtain
 the desired generating function
\be
 Z_{\rm CBC} = \sqrt {\prod_{n=0}^{N-1}  \detB_M(\theta_n)}, \quad\quad \theta_n=(2n+1)\pi/N.\nn
\ee

For symmetric weights $x'=x,y'=y,z'=z$ we have
\bea
 a & = & 8 x^2 y^2 z^2(1-\cos\theta), \quad
 b = 8\im x y^3 z^3 \sin\theta\nn \\
 c & = & -8\im x^3 y  z \sin\theta, \quad\quad\quad
 d = 8 x^2 y^2 z^2 (1+\cos\theta) \nn \\
 \lambda_+ &=&16 x^2y^2z^2, \quad\quad \quad\quad \  \lambda_- =0.\nn
\eea
 We obtain
  $\detB_M(\theta)=(16x^2y^2z^2)^M \sin^2 (\theta/2)$
 and hence
\be
 Z_{\rm CBC} = (4xyz)^{MN} \prod_{n=0}^{N-1}\sin(\theta_n/2) = 2^{1-N} \cdot (4 x y z)^{MN}.
  \label{asym1}
\ee
 Again, we shall see in \sect{sec:spin} that this simple expression can be
 understood using a spin-variable mapping. 
 
 In the case of an infinite lattice and since $\lambda_+>\lambda_-$,
 we have from (\ref{detNM}) $\detB_M (\theta) \to \lambda_+^M$. 
 This gives the per-dimer free energy
\bea
 f&=& \lim _{M,N\to\infty} \frac 1 {3MN} \ln Z_{CBC}
  = \lim _{N\to \infty} \frac 1 {6N}\sum_{n=0}^{N-1}\ln \lambda_+(\theta_n)
  \nn\\
 &=& \frac 1 {12\pi} \int_0^{2\pi}\ln \lambda_+(\theta)\, \dif \theta 
 \label{freeCBC}
\eea
where $\lambda_+(\theta)$  is given by (\ref{lambda}). It is readily
verified that ({\ref{freeCBC}}) is identical to the free energy
(\ref{freePBC}) after carrying out the integration over $\phi$ in
(\ref{freePBC}).

\section{A spin-variable mapping}\label{sec:spin}
\begin{figure}
 \includegraphics{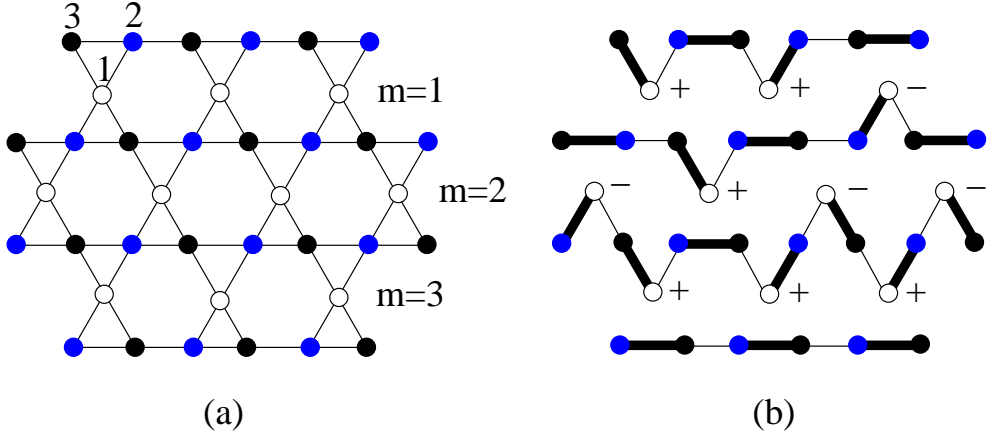}
\caption{ (a) An $\detM = 3$  kagome lattice. 
 Sublattice 1 sites are denoted by open circles. 
 (b) The spin configuration deduced from a typical dimer configuration.  
 The lattice is decomposed into 4 strips or loops depending on 
 the boundary condition (see text).  
} \label{fig:spin}
\end{figure}
 The exact enumeration results (\ref{Asym}) and (\ref{asym1})
 for a lattice of $M\times N$ unit cells with {symmetric} dimer weights
 are strikingly simple, suggesting the possibility of
 a simple derivation. 
 Indeed, Zeng and Elser \cite{ElserZeng} and Misguich {\it et al.} \cite{MSP}
 have introduced a pseudo-spin consideration of  enumerating quantum states 
 which can be transcribed to 
the present   $x=y=z=1$ case \cite{Misg}.
 However, the pseudo-spin consideration was presented in contexts of 
 spin 1/2 antiferromagnets and quantum dimer models, and the
 application to the classical dimer problem
 with general weights $x,y,z$ is not immediately obvious. 
 A simpler formulation is very much needed.

 We elucidate the matter by describing an alternate spin-variable mapping 
 valid for general $x,y,z$.   
 First we note  that the numbers of $x, y,$ and $z$ dimers
 are always fixed for finite lattices. This is due to
 the fact that the three principal axes
 do not intersect at  common
 points.  The kagome lattice has three sublattices as numbered in \fig{fig:spin}(a).
 Denote the number of sites on sublattice $i$
 by $N_i$, $i=1,2,3$, and  the number
 of $x$ dimers by $N_x$, etc. Then as  a consequence of the fact
 the principal axes do not intersect at common points, we have the relations
\be
 N_y + N_z = N_1, \quad
 N_z + N_x = N_2, \quad
 N_x + N_y = N_3\, .\nn   \nn
\ee
 This gives rise to $N_x = (N_2+N_3-N_1)/2$, etc., which are fixed numbers.
 The dimer generating function is therefore a single monomial of the form
\be
 Z= \Omega \, x^{N_x}y^{N_y}z^{N_z},\nonumber
\ee
 so we need only to compute the  constant $\Omega$.
In the case of $N_1=N_2=N_3={\cal N}$ we are considering, this leads to the  expression
\be
Z= \Omega\, (xyz)^{{\cal N}/2}, \label{Omega}
\ee
where we have
${\cal N} = 2MN$ for both the PBC and CBC boundary conditions.
 
We next  map dimer configurations on the lattice to
 spin configurations on one sublattice,
 say, 1.  
  Consider the example of the lattice shown in \fig{fig:spin}(a). 
 Let the lattice consist $\detM$ rows of  sublattice $1$ sites
 and $\detM+1$ rows of equal $2$ and  $3$ sites with open boundaries
 in the vertical direction.
 The boundary condition in the horizontal direction
 can be either open or periodic.
 Denote the number of sublattice 1 sites in the $m$th  row by  $n_m$, which must
 satisfy  the sum rule
\be
 n_1+n_2+\cdots + n_{\detM}= N_1= {\rm even}, \label{sumrule1}
\ee
 since the lattice must admit dimer
 coverings.

 Assign spin variable
\be
 \sigma_{mn} =\pm 1,\quad\quad m=1, ..., \detM, \ n=1, ..., n_m \nn
\ee
 to sublattice 1 sites, where  $m$ is the row number counting beginning from the top.
 Adopt the  convention  that $\sigma=+1$ $(-1)$ 
 if the dimer covering the site also covers a site above (below) the row. 
 For $+1$ ($-1$) sites we  remove the two edges below (above) the site 
 as well as the horizontal edge directly above (below) it 
 as shown in \fig{fig:spin}(b).
 This procedure decomposes the lattice into  strips (loops)
 for open (periodic) boundary conditions in the horizontal direction.

 Now every strip (or loop) must
 have an even number of sites to accommodate one  (or 2) dimer
 covering(s).  This condition imposes constraints on  spin configurations that can be
 realized by this mapping.  To best describe the constraints it
 is convenient to define a row variable
\be
 \tau_m = \prod_{n=1}^{n_m} \s_{mn}, \quad\quad m=1,2,...,\detM. \label{rowvariable}
\ee
 It is then readily verified that  we must have
\bea
 && \tau_1 = (-1)^{n_1}, \qquad \qquad  \nn \\
 &&\tau_{m-1}\tau_m = (-1)^{n_m},\ \quad m=2,3, ..., \detM\, , \nn\\
 && \tau_{\detM}=1.
\label{constraint1}
\eea
 The constraint $\tau_{\detM}=1$ is automatically satisfied due
 to (\ref{sumrule1}) and the fact that
\be
 \tau_M = (\tau_1)(\tau_1\tau_2) \cdots (\tau_{M-1}\tau_M) 
= (-1)^{m_1 \,+ \,\cdots \,+ \ m_M} = (-1)^{N_1} =1. 
\ee
 We remark that the row variable (\ref{rowvariable}) can also be used to
 analyze the dimer models in higher dimensions considered in \cite{Dhar}.

 We can now compute the constant $\Omega$. 
 Beginning with an overall  spin state degeneracy  $2^{N_1}$ of 
 sublattice 1 sites, each constraint in (\ref{constraint1}) reduces 
 the spin states by a factor of 2. 
 Since there are $\detM - 1$ such constraints, we have
 \be
\Omega=\left \{\begin{array}{ll}
   2^{N_1-(\detM-1)}, & {\rm open\ boundaries}\\
   2^{N_1+2}, & {\rm horizontal\ periodic\ boundary\ condition}
\end{array}\right.\label{GenCounting}
\ee
Note that there is an extra factor $2^{\detM+1}$ for periodic boundary conditions
in the horizontal direction since each loop
  has 2 dimer coverings. 
  Expression (\ref{GenCounting})  is a very general result independent of 
  specific values of $n_m$.

  For a lattice of $M\times N$ unit cells with toroidal boundary
  conditions PBC considered in \sect{sec:periodic}, we have $N_1=N_2=N_3 ={\cal N} =  2MN$,
  $\detM = 2M$.  Hence (\ref{GenCounting})  gives
  $\Omega = 2^{N_1}\cdot 2^{-(\detM-1)} \cdot 2^{\detM}$,
  where as in (\ref{GenCounting}) the second  factor is due to $\detM-1$
  constraints with the $\detM$th constraint
  automatically satisfied, and the third
  factor is due to the 2-fold dimer coverings of each loop.
  This leads to  $Z_{\rm PBC} = 2\cdot (4xyz)^{MN}$ in agreement
  with (\ref{Asym}).

 For a lattice of $M\times N$ unit cells with cylindrical
 boundary condition CBC considered in \sect{sec:cylindrical}, we again
 have $N_1=N_2=N_3 = {\cal N} = 2MN$,
  $\detM = 2M$.  However, the $\detM$th row of $N$
  sublattice 1 spins
  must be all $+1$ (Cf. \fig{fig:spin}(b) with the bottom
  row of sites removed) reducing the counting by a factor of $2^{-N}$ and the number of
rows by 1.
Hence $\Omega = 2^{-N} \cdot 2^{N_1 -(\detM -1 )} \cdot  2^{\detM }$ and
  $Z_{\rm CBC} = 2^{1-N} \cdot (4xyz)^{MN}$ in agreement
  with (\ref{asym1}).

 We remark that our spin-variable mapping  is akin to 
 one used recently by
 Dhar and Chandra \cite{Dhar}. However,  
  the Dhar-Chandra approach focuses  on an infinite lattice by
   ignoring what happens on the boundary.  Here, 
  we treat the boundary effect rigorously and   apply the mapping to finite lattices.

\ack
We are grateful to D. Dhar for sending a copy of
\cite{Dhar} and to G. Misguich for calling
our attention to  \cite{ElserZeng}  
- \cite{Misg}.
The work by FW is supported in part by grant LBNL DOE-504108.

\newpage

\setcounter{equation}{0}
\setcounter{section}{0}
\setcounter{figure}{0}

\begin{frontmatter}
\title{Dimers on the kagome lattice II: \\
Correlations and the Grassmannian approach}

\author[Berkeley1,LBL1]{Fa Wang},
\author[Boston1]{F. Y. Wu}
\address[Berkeley1]{ Department of Physics, University of California, Berkeley, California 94720 }
\address[LBL1]{ Material Sciences Division, Lawrence Berkeley National Laboratory, Berkeley, California 94720 }
\address[Boston1]{ Department of Physics, Northeastern University, Boston, Massachusetts 02115 }

\date{Printed \today}

\begin{abstract}
In this paper we continue our consideration of closed-packed dimers on the
kagome lattice.  Using the Pfaffian approach we evaluate the correlation between dimers on two
lattice edges.  It is found that the correlation is extremely short-ranged
in the case of symmetric dimers weights. Explicit expressions for the nonvanishing
correlations  are obtained in the interior of a large lattice.
 We also describe  a Grassmannian functional integral
approach, and use it to evaluate the dimer generating function and correlation functions.

\end{abstract}

\begin{keyword}
kagome lattice, close-packed dimers, correlation function, Grassmannian approach
\PACS 05.50.+q \sep 04.20.Jb \sep 02.10.Ox
\end{keyword}

\end{frontmatter}

\section{Introduction}\label{sec:intro}
in the preceeding paper  \cite{paperI}, hereafter referred to as \paperI,
 we presented exact results on 
the generating function for closed-packed dimers on a finite kagome lattice
with asymmetric dimer weights.
  To further illustrate the usefulness of the
Pfaffian method used in \paperI, in this paper we extend the consideration to dimer-dimer 
correlations. 
 For symmetric dimer
weights we find that the correlation is extremely short-ranged, a property unique to
the kagome lattice and previously reported by us in \cite{WangWu1}. Here we derive
explicit expressions of nonvanishing correlation functions for
 a large lattice.
We also describe the formulation of a Grassmannian function integral
approach, and use it to evaluate the dimer generating funciton and correlation
functions.

\section{The dimer-dimer correlation function}\label{sec:dimercorrelation}
The dimer-dimer correlation function measures the correlation between 
dimers on two lattice edges. This correlation is best described by introducing
  an edge occupation number
 \be
 n_{ij}=\left \{\begin{array}{rl}
1 & {\rm if\>edge}\>ij{\rm\ is\ occupied\  by \ a \ dimer} \\
0 & {\rm if\ edge}\>ij{\rm\ is\ empty} .       \label{equ:edgecovering} \\
\end{array}
\right.
 \ee
Likewise, the edge vacancy number is $\bar{n}_{ij}=1-n_{ij}$. 
The
correlation function between two dimers covering edge $ij$ in unit
cell at  $\vecr_1$ and edge $k\ell$ in unit cell at  $\vecr_2$ is defined by
 \be
\begin{split}
c(ij,\vecr_1;k\ell,\vecr_2)
& = \bra n_{ij,\vecr_1}\,n_{k\ell,\vecr_2}\ket - \bra n_{ij,\vecr_1}\ket \bra n_{k\ell,\vecr_2}\ket \\
& = \bra \bar{n}_{ij,\vecr_1}\,\bar{n}_{k\ell,\vecr_2}\ket - 
\bra \bar{n}_{ij,\vecr_1}\ket \bra \bar{n}_{k\ell,\vecr_2}\ket 
\end{split}
\label{equ:correlation}
\ee
where $\bra\cdot\ket$ denotes the configuration average.

The second line in (\ref{equ:correlation}) is useful in  computing 
the correlation function in the Pfaffian approach \cite{Fisher},
since using it we
 need only to keep track of the dimer generating function with specific
edge(s) missing as dictated by $\bra\bar{n}\ket$ or
$\bra\bar{n}\bar{n}\ket$.  

Let $A$ be the
antisymmetric Kasteleyn matrix derived from a Kasteleyn orientation, and
let $A'$ denote the antisymmetric matrix derived from $A$ with
edge $ij$, say in computing $\bra \bar{n}_{ij} \ket$, missing.
Write
 \be
Z  = \pfa A, \quad Z' = \pfa A' = \pfa [A+\Delta]
 \ee
  where $\Delta$ is the
matrix with zero elements everywhere except the $ij$ element is
$-A_{ij}$ and the $ji$ element is $-A_{ji}$. Then
 \be
 \bra \bar{n}_{ij} \ket = {Z'}/ Z = \pfa A' / \pfa A
 \nn \ee
 and
 \be \bra
\bar{n}_{ij} \ket ^2 = \frac{\det A'}{\det A}
 = \frac{ \det[ A (\idmat + G \Delta)]}{\det A}
 = \det (\idmat+G \Delta ).
   \label{equ:gf}
\ee
where $ G \equiv A^{-1}$ is the Green's function matrix and $\idmat$ the identity matrix.

In (\ref{equ:gf}) we need only to keep those row(s) and column(s)
in $\Delta$ and $A^{-1}$ where elements of $\Delta$ are nonzero.
This effectively reduces dimensions of matrices $G$ and $\Delta$ to 
 at most $4\times 4$ in the computation of (\ref{equ:correlation}). Explicitly, we have
 \bea
\bra \bar{n}_{ij,\vecr_1} \ket^2 & =& \det \bigg| \idmat_{2} 
+ \begin{pmatrix}
G(\vecr_1,\vecr_2)_{ii} & G(\vecr_1,\vecr_2)_{ij}\\
G(\vecr_1,\vecr_2)_{ji} & G(\vecr_1,\vecr_2)_{jj}
\end{pmatrix} 
\begin{pmatrix}
 0      & -A_{ij}\\
-A_{ji} &  0
\end{pmatrix} \bigg| \nn \\ 
\bra \bar{n}_{ij,\vecr_1}\,\bar{n}_{k\ell,\vecr_2}\ket^2 & =&
  \det | \idmat_{4} + \tilde{G}\tilde{\Delta} |,
 \eea
where $\idmat_{n}$ is the $ n\times n$ identity matrix, $\tilde{G}$ and $\tilde{\Delta}$ 
are the $4\times 4$ matrices
 \bea
\tilde{G} &=&\begin{pmatrix}
G(\vecr_1,\vecr_1)_{ii}     & G(\vecr_1,\vecr_1)_{ij}     & G(\vecr_1,\vecr_2)_{ik}     & G(\vecr_1,\vecr_2)_{i\ell}\\
G(\vecr_1,\vecr_1)_{ji}     & G(\vecr_1,\vecr_1)_{jj}     & G(\vecr_1,\vecr_2)_{jk}     & G(\vecr_1,\vecr_2)_{j\ell}\\
G(\vecr_2,\vecr_1)_{ki}     & G(\vecr_2,\vecr_1)_{kj}     & G(\vecr_2,\vecr_2)_{kk}     & G(\vecr_2,\vecr_2)_{k\ell}\\
G(\vecr_2,\vecr_1)_{\ell i} & G(\vecr_2,\vecr_1)_{\ell j} & G(\vecr_2,\vecr_2)_{\ell k} & G(\vecr_2,\vecr_2)_{\ell\ell}\\
\end{pmatrix} , \nn \\
\tilde{\Delta} &=& \begin{pmatrix}
 0      & -A_{ij} &  0          &  0\\
-A_{ji} &  0      &  0          &  0\\
 0      &  0      &  0          & -A_{k\ell}\\
 0      &  0      & -A_{\ell k} &  0
\end{pmatrix} .  \nn
 \eea
The formulation so far is very general applicable to any finite lattice with
symmetric or asymmetric dimer weights and
$\vecr_1$ and $\vecr_2$ arbitrary.
We now specialize  to a large lattice with symmetric dimer weights.
  
In the interior of a large lattice, the
correlation depends only on the difference $\vecr = \vecr_1 -
\vecr_2 = \{r_x, r_y\}$, so elements of $G$ are given by
 \be
G(\{r_{1x},r_{1y}\};\{r_{2x},r_{2y}\})_{ij} =  
\int_{0}^{2\pi}{\int_{0}^{2\pi}{}} \frac{\dif \theta \,\dif \ph}{(2\pi)^2}\,
e^{\im[(r_{1x}-r_{2x})\theta +(r_{1y}-r_{2y})\ph]}
A^{-1}(\theta,\ph)_{ij}. \label{equ:greensfunction}
 \ee
 For symmetric weights, the   $6\times 6$ inverse matrix $A^{-1}(\theta,\ph)$ in
(\ref{equ:greensfunction}) is computed by using  Equ.~(7) in {\paperI}, yielding
 \be
A^{-1}(\theta,\ph) = \frac{1}{4 x y z} 
 \begin{pmatrix}
 P_{3\times 3}+Q_{3\times 3}  &  R_{3\times 3} \\
-R^\dagger_{3\times 3}        &  P^\nd_{3\times 3}-Q^\nd_{3\times 3}
\end{pmatrix}, \label{equ:Ainverse}
\ee
where
\be
 P_{3\times 3}=\begin{pmatrix}
0                   &  x y e^{-\im\theta} &  x z \\
-y x e^{\im\theta}  &  0                  &  -y z e^{\im\theta} \\
-z x                &  z y e^{-\im\theta} &  0
 \end{pmatrix}, \quad
 Q_{3\times 3}=\begin{pmatrix}
x^2 (e^{-\im\theta}-e^{\im\theta}) &  -x y  &  -x z e^{\im\theta} \\
y x                                &  0     &  -y z \\
z x e^{-\im\theta}                 &  z y   &  0
\end{pmatrix}, \nn
\ee
\be
 R_{3\times 3}=\begin{pmatrix}
0 & -x y (e^{-\im(\theta+\ph)}+e^{-\im\ph}) & x z (e^{-\im(\theta+\ph)}+e^{-\im\ph}) \\
y x (e^{\im\theta}-1) & y^2 (1-e^{-\im\ph}) & y z (e^{-\im\ph}+e^{\im\theta}) \\
z x (e^{\im\theta}-1) & z y (1-e^{-\im(\theta+\ph)}) & z^2 (e^{-\im(\theta+\ph)}+e^{\im\theta})
\end{pmatrix} \nn
 \ee
 and the superscript $\dagger$ denotes Hermitian conjugation.

  Substituting (\ref{equ:Ainverse}) into
(\ref{equ:greensfunction}), we find
 $G(\{r_{1x},r_{1y}\};\{r_{2x},r_{2y}\})_{ij}=0$ if
$|r_{1x}-r_{2x}|\geq 2$ or $|r_{1y}-r_{2y}|\geq 2$, or,  explicitly,
 \be
 c(ij, {\bf r}_1;k\ell, {\bf r}_2) =0, \quad |{\bf r}_1- {\bf r}_2| \geq
 2.\label{zerocorrelation}
  \ee
  Equation  (\ref{zerocorrelation}) says that the correlation function vanishes identically
if the distance between the two lattice edges under consideration
 is larger than two lattice spacing.  This is
a consequence of the simple form of the
free energy given by Equ.~(13) in \paperI.
As  pointed out in \cite{WangWu1}, the absence of the
dimer-dimer correlation beyond a certain distance,  also
found in the Sutherland-Rokhsar-Kivelson state of a quantum dimer
model \cite{Misguich,Rokhsar}, is a property unique to the kagome lattice.

\begin{figure}
 \includegraphics{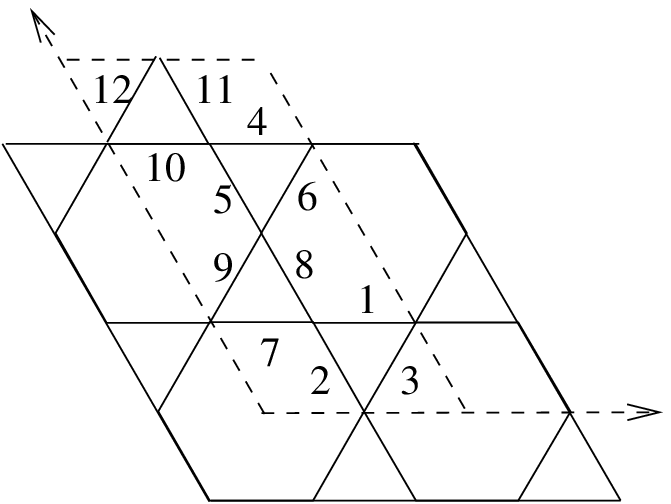}
 \caption{Labeling of the twelve edges belonging to one unit
cell.} \label{fig:edges}
\end{figure}
We can now compute all nonvanishing correlation functions
by substituting ({\ref{equ:greensfunction}) and (\ref{equ:Ainverse}) 
into (\ref{equ:correlation}), including
correlations
between two dimer edges within a unit cell and in two neighboring unit
cells. There are 12 edges belonging to a unit cell as numbered in
\fig{fig:edges}.  Details of the computations, which are straightforward,
will not be given. We
tabulate the  results   in Tables~\ref{Table1}-\ref{Table3}.

\section{Grassmannian approach}\label{sec:grassmannian}
In a series of papers in 1980 \cite{Samuel} Samuel pointed out and explored the
relation between the classical close-packed dimer problem and the
free Grassmannian lattice field theory. 
He showed that the dimer generating function can be represented as 
a fermionic path integral, and that
dimer correlation functions are 
 correlators occurring in a free fermion theory. 
He applied this formalism to dimers on the square lattice.
 Subsequently, Fendley {\it et al.} \cite{Fendley} extended the consideration
to the triangular 
lattice using a slightly modified fermionic action; they also elucidated 
the field theory aspect of the formalism. 
Here we follow the formalism of 
\cite{Fendley} and apply it to the kagome lattice.

We begin with the Kasteleyn orientation of the lattice shown in Fig.~1 in {\paperI}.
Associate  {\em real} Grassmannian variables
 $\eta$ to  lattice sites and consider the functional integral
 \be
 Z=\int{\exp\Big[(1/2)\sum_{i,j}{\eta_i A_{i j} \eta_j}\Big] \mathcal{D}\eta}\, ,
 \label{equ:pathintegral1}
 \ee
where $A_{ij}$ is the matrix element 
of the $6MN\times 6MN$ Kasteleyn matrix $A$ given by Equ.~(1) in {\paperI} and 
$\mathcal{D}\eta = \prod_i \dif \eta_i$. 

It is well-known that the integral (\ref{equ:pathintegral1}) is the Pfaffian
of the matrix $A$, a fact
 which 
 can be seen by expanding the exponential  and using the result that the integral 
 \be
 \int{\eta_{i_1}\dots\eta_{i_{6MN}}\mathcal{D}\eta}=\pm 1, \nn
 \ee
if $i_1\cdots i_{ 6MN} $ is a permutation of $1,\cdots,6MN$, and vanishes otherwise.  
 Thus, the  path integral (\ref{equ:pathintegral1}) is identical to the dimer 
generating function. 
 
For finite lattices with toroidal boundary conditions PBC considered in  Section~2.1 in {\paperI},
the dimer generating function is again expressed as a linear combination 
of four integrals corresponding to periodic boundary conditions with, or without,
the reversal of boundary edge orientations as discussed in Section 2.1 in \paperI.            .
 In each case the exponent 
 in (\ref{equ:pathintegral1}) 
can be block-diagonalize by  Fourier transforms of $\eta_i$. 

For reverse edge orientations in both directions, for example, 
the Fourier transform is
 \be
 \eta_{\alpha_p,\beta_q,a}=(MN)^{-1/2}\sum_{n,m}{\eta_{n,m,a} e^{\im \alpha_p n+\im \beta_q m}},\nn
 \ee
where $(n,m)$ labels the position of unit cells,
$a=1, \dots, 6$ the sites within a unit cell, 
and $\alpha_p=(2p+1)\pi/N,\ p=0,\dots ,N-1$, $\beta_q=(2q+1)\pi/M,\ q=0,\dots ,M-1$.
 
After carrying out the Fourier transform,
the functional integral (\ref{equ:pathintegral1}) becomes
 \be
 Z=\int{\exp\Big\{\frac 1 2\sum_{p,q,a,b}{ \eta_{\alpha_p,\beta_q,a}^*
 \Big[A(\alpha_p,\beta_q)\Big]_{ab} \eta_{\alpha_p,\beta_q,b} } \Big\} \mathcal{D}\eta},
 \label{equ:pathintegral2}
 \ee
where $A(\alpha,\beta)$ is the $6\times 6$ matrix 
given in Equ.~(7) in paper {\paperI} and
$\mathcal{D\,\eta }= \prod_{p,q,a}{\dif \eta_{\alpha_p,\beta_q,a}  }$. 
 Further using relations
\be
\eta_{\theta_p,\ph_q,a}^* = \eta_{\alpha_{N-1-p},\beta_{M-1-q},a},  \nn
\ee
and
\be
\big[A(\alpha_p,\beta_q)\big]_{a,b} = \big[-A(-\alpha_p,-\beta_q)\big]_{ba} = \big[-A(\alpha_{N-1-p},\beta_{M-1-q})\big]_{ba},
\ee
 we obtain from (\ref{equ:pathintegral2})
 \be
 \begin{split}
 Z& =\int{e^{(1/2)\sum_{p,q,a,b}{ \eta_{\alpha_{N+1-p},\beta_{M-1-q},a} 
  \left[-A(\alpha_{N-1-p},\beta_{M-1-q})\right]_{ba} \eta_{\alpha_{N-1-p},\beta_{M-1-q},b}^* } 
  } \mathcal{D}\eta^*}\\
  & =\int{e^{(1/2)\sum_{p,q,a,b}{ \eta_{\alpha_{N+1-p},\beta_{M-1-q},b}^* 
  \left[A(\alpha_{N-1-p},\beta_{M-1-q})\right]_{ba} \eta_{\alpha_{N-1-p},\beta_{M-1-q},a} } } \mathcal{D}\eta^*}\\
  & =\int{e^{(1/2)\sum_{p,q,a,b}{ \eta_{\alpha_{p},\beta_{q},a}^* \left[A(\alpha_{p},
    \beta_{q})\right]_{ab} \eta_{\alpha_{p},\beta_{q},b} } } \mathcal{D}\eta^*}\label{equ:pathintegral3}
\end{split}
 \ee
where $\mathcal{D}\eta^*=\prod_{p,q,a}{\dif \eta_{\alpha_p,\beta_q,a}^* }=
\prod_{p,q,a}{\dif \eta_{\alpha_{N-1-p},\beta_{M-1-q},a} }=\mathcal{D}\eta$.  Here, 
 we have used anticommutation relations of Grassmannian variables 
and renamed dummy variables $p,q,a,b$.

Next we use a well-known formula of Gaussian integrals of Grassmannian variables 
(see, e.g., (2.8) in the first reference in \cite{Samuel}) to write the product of 
(\ref{equ:pathintegral2}) and (\ref{equ:pathintegral3}) as
 \be
 Z^2 = \prod_{p,q}{ \iint{ e^{\sum_{a,b}{ \eta_{\alpha_p,\beta_q,a}^* [A(\alpha_p,\beta_q)]_{ab}
 \eta_{\alpha_p,\beta_q,b} } } \mathcal{D}\eta_{p,q}\mathcal{D}\eta_{p,q}^*
  } } = \prod_{p,q}{\det|A(\alpha_p,\beta_q)|}. \nn
 \ee
where $\mathcal{D}\eta_{p,q}\mathcal{D}\eta_{p,q}^*=\prod_{a}{\dif \eta_{\alpha_p,\beta_q,a} \dif \eta_{\alpha_p,\beta_q,a}^*}$. 
This is identically the expression $\det |A_4|$ given by Equ.~(6) in {\paperI} 
derived there using the method of Pfaffians.

The Grassmannian formalism is most useful in evaluating
dimer-dimer correlation functions.
To evaluate the correlation function 
(\ref{equ:correlation}), for example, we note that
the dimer generating function can be written as 
 \be
 Z=\sum_{\rm covering}{\Big[\prod_{<ij>}w_{ij}^{n_{ij}}\Big] }\, ,
 \nn \ee
where $n_{ij}$ is defined in (\ref{equ:edgecovering}) and
the summation  is over all dimer coverings of the lattice.
Thus, we have
 \be
 \begin{split}
 \bra n_{ij}\ket &= w_{i j}
  Z^{-1} \frac{\partial Z}{\partial w_{ij} }  = w_{i j} Z^{-1} \int{\eta_i\eta_j 
  \exp ({ \mathcal{A} }) \mathcal{D}\eta}, \\
  \bra n_{ij}n_{k\ell}\ket &=  w_{i j} w_{k \ell}
 Z^{-1} \frac{\partial^2 Z}{\partial w_{k\ell}\partial w_{ij} } 
 = w_{i j} w_{k \ell} Z^{-1}\int{\eta_i\eta_j\eta_k\eta_\ell \exp({\mathcal{A}})  
\mathcal{D}\eta}. 
 \end{split}
 \nn 
 \ee
where $\mathcal{A}=(1/2)\sum_{i,j}{\eta_i A_{i j} \eta_j}$, and 
we have assumed that the Kasteleyn orientation is from $i$ to $j$ and 
from $k$ to $\ell$ to get the correct sign for the Grassmannian integrals.  
 These are precisely expressions of correlators occurring
in a Gaussian theory and can be immediately written down.  This gives
 \bea
 \bra n_{ij}\ket & = & w_{i j} A^{-1}_{j i} \nn \\
 \bra n_{ij}n_{k\ell}\ket & = &  w_{i j} w_{k \ell}(A^{-1}_{j i}A^{-1}_{\ell k} 
  - A^{-1}_{k i}A^{-1}_{\ell j} + A^{-1}_{\ell i}A^{-1}_{k j}), \nn
 \eea 
where $A^{-1}$ is the inverse matrix of $A$ with elements given by Equ.~(1) in {\paperI}. 
Here, we have used the Wick's theorem in expanding the
four-point correlator. 

Finally, the dimer correlation function (\ref{equ:correlation}) is given simply by
the expression
 \be
 c(ij;k\ell)= w_{i j} w_{k \ell}(
  - A^{-1}_{k i}A^{-1}_{\ell j} + A^{-1}_{\ell i}A^{-1}_{k j}),
\label{correla}
 \ee
where $A^{-1}\equiv G$ for a large lattice has been given  in
(\ref{equ:greensfunction}).  

It can be verified that (\ref{correla}) gives rise to the same results 
as tabulated in Tables~\ref{Table1}-\ref{Table3}.
Correlation functions of three or more dimers can be derived 
in a similar fashion. The advantage of using the Grassmannian method here
is that it does not  invoke products of large matrices needed in the Pfaffian approach.
  
\ack
 The work by FW is supported in part by grant LBNL DOE-504108.

\newpage
\begin{table}
\begin{tabular}{r|rrrrrrrrrrrr}
\hline\hline
\multicolumn{13}{c}{$(r_x,r_y)=(0,0)$}\\
\hline
   &  1 &  2 &  3 &  4 &  5 &  6 &  7 &  8 &  9 & 10 & 11 & 12 \\
\hline
 1  & \id& \m & \m &  0 &  0 &  0 & \m & \m & \p &  0 &  0 &  0 \\
 2  & \m & \id& \m &  0 &  0 &  0 & \m & \m & \p &  0 &  0 &  0 \\
 3  & \m & \m & \id&  0 &  0 &  0 & \p & \p & \m &  0 &  0 &  0 \\
 4  &  0 &  0 &  0 & \id& \m & \m & \m & \p & \p & \m & \m & \p \\
 5  &  0 &  0 &  0 & \m & \id& \m & \p & \m & \m & \m & \m & \p \\
 6  &  0 &  0 &  0 & \m & \m & \id& \p & \m & \m & \p & \p & \m \\
 7  & \m & \m & \p & \m & \p & \p & \id& \m & \m &  0 &  0 &  0 \\
 8  & \m & \m & \p & \p & \m & \m & \m & \id& \m &  0 &  0 &  0 \\
 9  & \p & \p & \m & \p & \m & \m & \m & \m & \id&  0 &  0 &  0 \\
10  &  0 &  0 &  0 & \m & \m & \p &  0 &  0 &  0 & \id& \m & \m \\
11  &  0 &  0 &  0 & \m & \m & \p &  0 &  0 &  0 & \m & \id& \m \\
12  &  0 &  0 &  0 & \p & \p & \m &  0 &  0 &  0 & \m & \m & \id\\
\hline\hline
\end{tabular}
\caption{Correlation function (\ref{equ:correlation}) between two dimers in the
 same unit cell, $a= 1/16, b=-1/16, s =3/16.$}
\label{Table1}
\end{table}

\begin{table}
\begin{tabular}{r|rrrrrrrrrrrr}
\hline\hline
\multicolumn{13}{c}{$(r_x,r_y)=(0,1)$}\\
\hline
 &  1 &  2 &  3 &  4 &  5 &  6 &  7 &  8 &  9 & 10 & 11 & 12 \\
\hline
 1  &  0 &  0 &  0 &  0 &  0 &  0 &  0 &  0 &  0 &  0 &  0 &  0 \\
 2  &  0 &  0 &  0 &  0 &  0 &  0 &  0 &  0 &  0 &  0 &  0 &  0 \\
 3  &  0 &  0 &  0 &  0 &  0 &  0 &  0 &  0 &  0 &  0 &  0 &  0 \\
 4  &  0 &  0 &  0 &  0 &  0 &  0 &  0 &  0 &  0 &  0 &  0 &  0 \\
 5  &  0 &  0 &  0 &  0 &  0 &  0 &  0 &  0 &  0 &  0 &  0 &  0 \\
 6  &  0 &  0 &  0 &  0 &  0 &  0 &  0 &  0 &  0 &  0 &  0 &  0 \\
 7  &  0 &  0 &  0 &  0 &  0 &  0 &  0 &  0 &  0 &  0 &  0 &  0 \\
 8  &  0 &  0 &  0 &  0 &  0 &  0 &  0 &  0 &  0 &  0 &  0 &  0 \\
 9  &  0 &  0 &  0 &  0 &  0 &  0 &  0 &  0 &  0 &  0 &  0 &  0 \\
10  & \m & \p & \p &  0 &  0 &  0 &  0 &  0 &  0 &  0 &  0 &  0 \\
11  & \p & \m & \m &  0 &  0 &  0 &  0 &  0 &  0 &  0 &  0 &  0 \\
12  & \p & \m & \m &  0 &  0 &  0 &  0 &  0 &  0 &  0 &  0 &  0 \\
\hline\hline
\end{tabular}
\caption{Correlation function (\ref{equ:correlation})  between two dimers in  unit cells
$(0,0)$ and $(0,1)$, $a= 1/16, b=-1/16.$
Row indices $1,2,\dots, 12$ label edges in unit cell $(0,0)$; 
column indices label  edges in  unit cell $(0,1)$ (Cf. \fig{fig:edges}).
} \label{Table2}
\end{table}

\begin{table}
\begin{tabular}{r|rrrrrrrrrrrr}
\hline\hline
\multicolumn{13}{c}{$(r_x,r_y)=(1,0)$}\\
\hline
 &  1 &  2 &  3 &  4 &  5 &  6 &  7 &  8 &  9 & 10 & 11 & 12 \\
\hline
 1  &  0 &  0 &  0 &  0 &  0 &  0 & \m & \p & \m &  0 &  0 &  0 \\
 2  &  0 &  0 &  0 &  0 &  0 &  0 & \p & \m & \p &  0 &  0 &  0 \\
 3  &  0 &  0 &  0 &  0 &  0 &  0 & \m & \p & \m &  0 &  0 &  0 \\
 4  &  0 &  0 &  0 &  0 &  0 &  0 &  0 &  0 &  0 & \m & \p & \m \\
 5  &  0 &  0 &  0 &  0 &  0 &  0 &  0 &  0 &  0 & \p & \m & \p \\
 6  &  0 &  0 &  0 &  0 &  0 &  0 &  0 &  0 &  0 & \m & \p & \m \\
 7  &  0 &  0 &  0 &  0 &  0 &  0 &  0 &  0 &  0 &  0 &  0 &  0 \\
 8  &  0 &  0 &  0 &  0 &  0 &  0 &  0 &  0 &  0 &  0 &  0 &  0 \\
 9  &  0 &  0 &  0 &  0 &  0 &  0 &  0 &  0 &  0 &  0 &  0 &  0 \\
10  &  0 &  0 &  0 &  0 &  0 &  0 &  0 &  0 &  0 &  0 &  0 &  0 \\
11  &  0 &  0 &  0 &  0 &  0 &  0 &  0 &  0 &  0 &  0 &  0 &  0 \\
12  &  0 &  0 &  0 &  0 &  0 &  0 &  0 &  0 &  0 &  0 &  0 &  0 \\
\hline\hline
\end{tabular}
\caption{Correlation function (\ref{equ:correlation})  between two dimers in  unit cells
$(0,0)$ and $(1,0)$, $a= 1/16, b=-1/16.$
Row indices $1,2,\dots ,12$) label  edges in unit cell $(0,0)$; 
column indices label  edges in the unit cell $(1,0)$ (Cf. \fig{fig:edges}).
} \label{Table3}
\end{table}

 \end{document}